\documentclass{jltp}

\usepackage{graphicx} % uncomment this line to include the graphicx package

\makeatletter
\renewenvironment{thebibliography}[1]
     {\section*{\MakeUppercase\refname
        \@mkboth{\MakeUppercase\refname}{\MakeUppercase\refname}}%
      \small\list{\@arabic\c@enumiv .}  %No brackets around ref. number, small font
           {\settowidth\labelwidth{#1.}  %No brackets around ref. number
            \leftmargin\labelwidth
            \setlength{\parsep}{0\p@ \@plus \p@} %Don't want space between references
            \setlength{\itemsep}{0\p@ \@plus \p@} %Don't want space between references
            \advance\leftmargin\labelsep
            \@openbib@code
            \usecounter{enumiv}%
            \let\p@enumiv\@empty
            \renewcommand\theenumiv{\@arabic\c@enumiv}}%
      \sloppy\clubpenalty4000\widowpenalty4000%
      \sfcode`\.\@m}
     {\def\@noitemerr
       {\@latex@warning{Empty `thebibliography' environment}}%
      \endlist}
\makeatother

\newcommand{\I}{{\rm i}}

\title{Mott--Hubbard Insulator in Infinite 
Dimensions\footnote{Dedicated to Peter W\"olfle on the occasion
of his 60th birthday.}}

\author{Eva Kalinowski and Florian Gebhard}

\address{Fachbereich Physik, Philipps-Universit\"at Marburg,
D-35032 Marburg, Germany}

\runninghead{E.~Kalinowski and F.~Gebhard}{Mott--Hubbard Insulator 
in Infinite Dimensions}

\begin{document}

\maketitle%\today

\begin{abstract}

We calculate the one-particle density of states for the Mott--Hubbard
insulating phase of the Hubbard model on a Bethe lattice
in the limit of infinite coordination number. We employ the
Kato--Takahashi perturbation theory around the strong-coupling limit
to derive the Green function.
We show that the Green function for the lower Hubbard band can be 
expressed in terms of polynomials in the bare hole-hopping operator.
We check our technique against the exact solution of the
Falicov--Kimball model and give explicit results up to and
including second order in the inverse Hubbard interaction. 
Our results provide a stringent test for analytical and numerical
investigations of the Mott--Hubbard insulator and the
Mott--Hubbard transition within the dynamical mean-field theory.
We find that the Hubbard-III approximation is not satisfactory beyond
lowest order, but the local-moment approach provides a very good
description of the Mott--Hubbard insulator at strong coupling.

\end{abstract}

\section{INTRODUCTION}

As emphasized by Mott early on\cite{Mott,Mottbook},
interacting electrons in a single half-filled band 
may undergo a quantum phase transition from a metallic to an insulating
state without breaking the translational
or spin symmetry of the underlying Hamiltonian
(Mott--Hubbard transition\cite{Gebhardbook}).
The explicit formulation of the corresponding minimal Hamiltonian by
Hubbard\cite{HubbardI}, Gutzwiller\cite{Gutzwiller}, and
Kanamori\cite{Kanamori}, allowed a concise discussion of the 
Mott--Hubbard transition. In the Hubbard model, the electrons'
on-site interaction of strength~$U$ 
competes with their kinetic energy,
characterized by the bandwidth~$W$ for vanishing interactions.

The Hubbard Hamiltonian
poses a most difficult many-body problem, and
approximate treatments had to be devised. For example,
within the Gutzwiller variational scheme\cite{Gutzwiller}
Brinkman and Rice\cite{BrinkmanRice} corroborated Mott's view of the
Mott--Hubbard transition as a {\sl continuous\/} quantum
phase transition: At some finite critical interaction strength, 
$U_{\rm c}={\cal O}(W)$, the
quasiparticle weight goes to zero coming from the
metallic side, and the gap closes coming from the insulating 
side\cite{Gebhardbook}.
Qualitatively the same picture is found from the exact solution of
the $1/r$~Hubbard model in one dimension\cite{GebhardRuckenstein,Gebhardbook}.

A few years ago, these views were challenged by Georges and Kotliar
and their collaborators; for a review, see Ref.~\onlinecite{RMP}.
They developed further and studied the dynamical mean-field
theory for the Hubbard model which becomes exact in the limit of infinite
lattice coordination number\cite{MVPRLdinfty}. Results from
various analytical and numerical techniques led them to the conclusion
that the Mott--Hubbard transition is {\sl discontinuous\/}
in the sense that the gap jumps to a finite value
when the quasiparticle weight becomes zero at $U_{{\rm c},2}\equiv U_{\rm c}$
(scenario of a `preformed gap'). They argue that 
the insulating solution to the 
equations persists down to $U_{{\rm c}, 1}<U_{{\rm c},2}$
but its energy is higher than that of the metallic state.

Various approaches to the Mott--Hubbard transition 
for lattices with infinite coordination number
yield results
in favor of either of the conflicting scenarios; for a review,
see Refs.~\onlinecite{Gebhardbook,RMP}, and 
%Refs.~\onlinecite{RDA,7Schwaben,7Schwabenreply,Krauth,%
%BullaNRGPRL,BullaVolli,BullaLDMFT} 
Refs.~11--17
for more recent treatments.
Insightful physical arguments helped to sharpen the 
physical and mathematical implications 
of the scenario of a preformed gap
%transition\cite{KotliarFisher,LoganNozieres,Kehrein,Kotliarreply} but cannot
transition$^{\mbox{\scriptsize 18--21}}$ but cannot
resolve the issue.
Since an exact solution to the dynamical mean-field equations is still
%lacking, all analytical approaches\cite{IPT,NCA,LMA,BullaLDMFT} 
lacking, all analytical approaches$^{\mbox{\scriptsize 22--24,17}}$
are necessarily approximate in nature.
Numerical investigations of the dynamical mean-field
%equations\cite{7Schwaben,7Schwabenreply,Krauth,BullaNRGPRL,BullaVolli,%
%BullaLDMFT,ED1,ED2}
equations$^{\mbox{\scriptsize 12--17,25,26}}$
involve (i)~discretization, (ii)~numerical diagonalization,
(iii)~interpolation, (iv)~iteration of the self-consistency cycle,
and (v)~extrapolation to the thermodynamic limit.
The Random-Dispersion Approximation\cite{RDA} provides
an alternative to the dynamical mean-field approach. However,
apart from~(iv), similar steps are required 
in its numerical implementation.
Therefore, the region of applicability of the available numerical
techniques is not a priori clear either.

In such a situation, benchmark tests need to be provided in order
to assess the
quality of the various analytical and numerical approaches deep
in the metallic and insulating phases where perturbation theory
provides reliable answers. If an
approximate technique fails in the perturbatively accessible
regimes where others may pass, its predictions on the
Mott--Hubbard {\sl transition\/} should not be taken too seriously.

In this work, we provide such a benchmark test for the Mott--Hubbard insulator
at zero temperature. In Sect.~\ref{SecII} we introduce
the definitions of the Hubbard model and
the Falicov--Kimball model\cite{Falicov} 
and the relevant one-particle Green functions.
We restrict ourselves to the half-filled Bethe lattice with
an infinite coordination number, and exclude all possible
ordered ground states which would conceal the Mott--Hubbard transition.

The ground state of the Mott--Hubbard insulator
is vastly degenerate.
Therefore, in Sect.~\ref{SecIII}, 
we invoke Kato--Takahashi perturbation theory
to calculate the one-particle Green function
systematically in the insulating regime. 
As a simple application, we re-derive 
the well-known lowest-order solution\cite{MVSchmit}.

In Sect.~\ref{SecIV} we use the exactly solvable Falicov--Kimball model as 
a test case\cite{PvD}. We show how to 
overcome the resummation problem
of our series expansion. We find excellent agreement
between our results to second order in $1/U$ and the exact density of states
down to $U\approx 0.75W$; for the Falicov--Kimball model,
$U_{\rm c}^{\rm FK}=W/2$ for the transition. 

This observation makes us confident that
our second-order results for the Hubbard model, Sect.~\ref{SecV},
provide a reliable test down to $U\approx 1.5~W$.
Our approach allows an estimate of the critical value for
the closure of the gap. This value corresponds to $U_{{\rm c},1}$
in the scenario of a discontinuous transition, and to $U_{\rm c}$
if the scenario of a continuous quantum phase transition is the correct
one; for the Hubbard model, there is consensus that the insulator
no longer exists below $0.9 W<U_{{\rm c},1}<1.3 W$.

In Sect.~\ref{SecVI} we make a first comparison with two
other analytical methods. The
Hubbard-III approximation turns out to be inappropriate whereas the
local-moment approach provides a very good description of the
Mott--Hubbard insulator.
A more detailed comparison is planned to be published elsewhere.
Short conclusions, Sect.~\ref{SecVII}, close our presentation.

\section{DEFINITIONS}
\label{SecII}

\subsection{Hamilton Operators}

In this work we study the Hubbard model\cite{HubbardI}, 
\begin{equation}
\hat{H}=\hat{T} + U \hat{D} \; .
\label{generalH}
\end{equation}
The kinetic energy of the electrons reads
\begin{equation}
\hat{T} = \frac{-t}{\sqrt{Z}} 
\sum_{i,\tau;\sigma} \hat{c}_{i,\sigma}^+\hat{c}_{i+\tau,\sigma} \; ,
\end{equation}
where $\hat{c}^+_{i,\sigma}$,
$\hat{c}_{i,\sigma}$ are creation and annihilation operators for
electrons with spin~$\sigma$ at site~$i$, 
where $i$~runs over all lattice sites and
$\tau$~denotes the $Z$~nearest-neighbor vectors.
Since we are interested
in the Mott insulating phase, we consider exclusively a half-filled band
where the number of electrons~$N$ equals the
number of lattice sites~$L$. 

We restrict our analysis to
the Bethe lattice where each site is connected to $Z$ other sites
without the generation of loops, and the limit $Z\to\infty$ is
implicitly understood henceforth. Then,
the non-interacting density of states
becomes\cite{Economou}
\begin{equation}
\rho(\omega)= \frac{2}{\pi W}\sqrt{4 -\left(\frac{4\omega}{W}\right)^2\,}
\quad , \quad   (|\omega|\leq W/2) \; ,
\label{rhozero}
\end{equation}
where $W=4t$ is the bandwidth. In the following, we shall set $t\equiv 1$
as our energy unit.

The Hubbard interaction reads
\begin{equation}
\hat{D} = \sum_{i} \left(\hat{n}_{i,\uparrow}-\frac{1}{2}\right)
\left(\hat{n}_{i,\downarrow}-\frac{1}{2}\right) \; ,
\end{equation}
where $\hat{n}_{i,\sigma}=
\hat{c}^+_{i,\sigma}\hat{c}_{i,\sigma}$ are the density operators for 
a given spin~$\sigma$.
For later use we also define the operators for the local electron density,
$\hat{n}_{l}=\hat{n}_{l,\uparrow}+\hat{n}_{l,\downarrow}$.
The operators $\hat{S}_i^x=(\hat{c}^+_{i,\uparrow}\hat{c}_{i,\downarrow}
+ \hat{c}^+_{i,\downarrow}\hat{c}_{i,\uparrow})/2$, 
$\hat{S}_i^y=(\hat{c}^+_{i,\uparrow}\hat{c}_{i,\downarrow}
- \hat{c}^+_{i,\downarrow}\hat{c}_{i,\uparrow})/(2\I)$, 
and $\hat{S}_i^z=(\hat{n}_{i,\uparrow}-\hat{n}_{i,\downarrow})/2$
are the three components of the spin-1/2 vector operator
$\hat{\vec{S}}_{i}$.

We have chosen the chemical potential in such a way 
that the Hamiltonian explicitly exhibits a particle-hole symmetry,
i.e., $\mu=0$ guarantees a half-filled band.
We restrict our analysis to the Mott--Hubbard insulator
at zero temperature which is characterized by a finite
ground-state entropy density\cite{RMP,Gebhardbook}, $s=\ln(2) + 
{\cal O}(\ln(L)/L)$. This means that in the Mott--Hubbard insulator
each lattice site is equally likely occupied by an electron with
spin $\uparrow$ or $\downarrow$, 
irrespective of the spin at any other lattice site.

Later we shall also address 
the case of the Falicov--Kimball model\cite{Falicov} 
(or simplified Hubbard model) in which only one species of electrons is mobile,
\begin{equation}
\hat{T}^{\rm FK} = \frac{-t}{\sqrt{Z}} 
\sum_{i,\tau} \hat{c}_{i}^+\hat{c}_{i+\tau} \; .
\label{FKkineticenergy}
\end{equation}
The mobile $c$~electrons interact with the immobile $f$~electrons 
via the Hubbard interaction,
\begin{equation}
\hat{D}^{\rm FK} = \sum_{i} \left(\hat{c}_{i}^+\hat{c}_i-\frac{1}{2}\right)
\left(\hat{f}_{i}^+\hat{f}_i-\frac{1}{2}\right) \; .
\end{equation}
The immobile electrons are {\sl randomly\/} distributed
over the lattice such that each lattice site is occupied by an $f$~electron
with probability $p=1/2$, irrespective of the $f$~occupation
at any other lattice site.

\subsection{Green Functions}

The time-dependent local single-particle Green function at zero temperature
is given by\cite{Fetter}
\begin{equation}
G(t) = -\I \frac{1}{L} \sum_{i,\sigma} 
\langle \hat{T} [ 
\hat{c}_{i,\sigma}(t)\hat{c}_{i,\sigma}^+] \rangle \; .
\label{GFdef}
\end{equation}
Here, $\hat{T}$ is the time-ordering operator, $\langle \ldots \rangle$
implies the average over all ground states with energy $E_0$,
and ($\hbar \equiv 1$)
\begin{equation}
\hat{c}_{i,\sigma}(t) = \exp(\I \hat{H} t) \hat{c}_{i,\sigma}
                          \exp(-\I \hat{H} t) 
\end{equation}
is the annihilation operator in the Heisenberg picture. 
Recall that the symmetries of~$\hat{H}$ are unbroken so that
each lattice site gives the same contribution in~(\ref{GFdef}).

The Bethe lattice is bipartite so that we may define
$(-1)^i=+1$ for the $A$~sites of the lattice which are
surrounded by $B$~sites only (and vice versa), for which $(-1)^i=-1$ 
($i\in B$).
Then, the particle-hole transformation 
\begin{equation}
{\cal T}: \qquad \hat{c}_{i,\sigma}^+ \mapsto (-1)^i
\hat{c}_{i,\sigma}\quad ; \quad
\hat{c}_{i,\sigma} \mapsto (-1)^i \hat{c}_{i,\sigma}^+
\end{equation}
maps $\hat{H}$ onto itself. 
We can readily identify the contributions from the lower (LHB)
and upper (UHB) Hubbard bands to the 
Fourier transform of the local Green function ($\eta=0^+$),
\begin{eqnarray}
G(\omega) &=& \int_{-\infty}^{\infty}dt e^{\I \omega t} G(t) 
= G_{\rm LHB}(\omega ) + G_{\rm UHB}(\omega ) \; , \nonumber \\
G_{\rm LHB}(\omega ) &=&  \frac{1}{L} \sum_{i,\sigma}
        \left\langle \hat{c}_{i,\sigma}^+
            \left[\omega +(\hat{H}-E_0)-\I\eta\right]^{-1} 
         \hat{c}_{i,\sigma}\right\rangle \; , \label{DefGLHB}\\
G_{\rm UHB}(\omega ) &=& - G_{\rm LHB}(-\omega )  \nonumber \; , 
\end{eqnarray}
due to particle-hole symmetry. 
Therefore, it is sufficient to
evaluate the local Green function for the lower Hubbard band
which describes the dynamics of a hole inserted into the ground state.

The density of states for the lower Hubbard band
can be obtained from the imaginary part of
the Green function~(\ref{DefGLHB}) for real arguments via\cite{Fetter}
\begin{equation}
D_{\rm LHB}(\omega) = \frac{1}{\pi} \Im\{ G_{\rm LHB}(\omega) \} 
\quad , \quad 
\mu_{\rm LHB}^- \leq \omega \leq   \mu_{\rm LHB}^+ <0
\; .
\label{rangeofD}
\end{equation}
Due to particle-hole symmetry we have 
$D_{\rm UHB}(\omega)=D_{\rm LHB}(-\omega)$. 
The gap in the Mott--Hubbard is symmetric around $\omega=0$ so that 
the one-particle gap in the Mott--Hubbard insulator is given by
\begin{equation}
\Delta = 2 | \mu_{\rm LHB}^+| >0 \; .
\label{Delta}
\end{equation}
For the Falicov--Kimball model we shall address
the local Green function of the mobile electrons for the lower Hubbard band 
\begin{equation}
G^{\rm FK}_{\rm LHB}(\omega ) =  \frac{1}{L} \sum_{i}
\left\langle \hat{c}_{i}^+
            \left[\omega +(\hat{H}-E_0)-\I\eta\right]^{-1} 
         \hat{c}_{i}\right\rangle \; .
\end{equation}
Again, the local Green function is given by
$G^{\rm FK}(\omega ) =  G^{\rm FK}_{\rm LHB}(\omega ) - 
G^{\rm FK}_{\rm LHB}(-\omega )$,
the density of states is symmetric around $\omega=0$, and 
Eq.~(\ref{rangeofD}) hold analogously.
For later reference we further define
\begin{equation}
\rho^{\rm FK}(\omega)= 
\frac{1}{2\pi}\sqrt{2 -\omega^2\,} \quad , \quad (|\omega|\leq \sqrt{2}) \; .
\label{rhoFKdef}
\end{equation}

\section{STRONG-COUPLING EXPANSION}
\label{SecIII}

\subsection{Kato--Takahashi Perturbation Theory}

Based on Kato's degenerate perturbation theory\cite{Kato},
Takahashi\cite{Takahashi} developed the perturbation
expansion in $1/U$ for the Hubbard model at zero temperature.
For large interaction strengths~$U$, 
the ground states $|\psi_n\rangle$
of $\hat{H}$ in~(\ref{generalH}) can be obtained from
states $|\phi_n\rangle$ without a double occupancy,
\begin{equation}
|\psi_n\rangle = \hat{\Gamma} |\phi_n\rangle \quad , \quad 
\hat{P}_0 |\phi_n\rangle =  |\phi_n\rangle \; ,
\label{gammaAction}
\end{equation}
where $\hat{P}_j$ projects onto the subspace with~$j$~double occupancies.

Eq.~(\ref{gammaAction}) is readily interpreted. In the large-coupling limit, 
$\hat{T}$ in~(\ref{generalH}) is considered as a perturbation on~$U\hat{D}$.
Addressing the ground states we therefore start from eigenstates 
with zero double occupancies~$|\phi_n\rangle$ into which
the operator $\Gamma$ successively introduces at most~$m$~double occupancies
to generate the ground states~$|\psi_n\rangle$ of~$\hat{H}$
to $m$-th order in~$t/U$.
The operator $\hat{\Gamma}$ reduces all operators
to the subspace with zero double occupancies. In particular,
$\hat{\Gamma}^+\hat{\Gamma}=\hat{P}_0$ so that overlap matrix elements 
obey\cite{Takahashi}
$\langle \psi_m | \psi_n\rangle = 
\langle \phi_m | \phi_n\rangle$.

The Schr\"odinger equation for the ground states
\begin{equation}
\hat{H}\hat{\Gamma} |\phi_n\rangle = E_0 \hat{\Gamma} |\phi_n\rangle 
\end{equation}
leads to
\begin{equation}
\hat{h}|\phi_n\rangle = 0 \quad , \quad 
\hat{h} = \hat{\Gamma}^+ \hat{H} \hat{\Gamma} -E_0 \hat{P}_0\; ,
\label{deflittleh}
\end{equation}
i.e., the eigenvalue does not change under the transformation.
The creation and annihilation operators
are transformed accordingly,
\begin{equation}
\hat{c}^+_{l,\sigma} \mapsto \widetilde{c}^+_{l,\sigma}=
\hat{\Gamma}^+  \hat{c}^+_{l,\sigma}
\hat{\Gamma} \quad , \quad 
\hat{c}_{l,\sigma} \mapsto \widetilde{c}_{l,\sigma}=
\hat{\Gamma}^+  \hat{c}_{l,\sigma}
\hat{\Gamma} \; .
\end{equation}
The derivation of the explicit expression for~$\hat{\Gamma}$
can be found in~Refs.~\onlinecite{Kato,Takahashi}, where it is
shown that
\begin{equation}
\hat{\Gamma} = 
\hat{P}\hat{P}_0 \left( \hat{P}_0\hat{P}\hat{P}_0 \right)^{-1/2} \; .
\label{DEFgamma}
\end{equation}
Here, 
\begin{eqnarray}
\hat{P} &=& \hat{P}_0 - \sum_{n=1}^{\infty} 
\sum_{{\scriptstyle r_1+\ldots+r_{n+1}=n}\atop {\scriptstyle r_i\geq 0}} 
\hat{S}^{r_1}\hat{T} \cdots \hat{T}\hat{S}^{r_{n+1}} \; , \nonumber \\
\hat{S}^0 &=& - \hat{P}^0 \; , \\
\hat{S}^r &=& \frac{(-1)^r}{U^r}\sum_{j\neq 0} 
\frac{\hat{P}_j}{j^r} \; .\nonumber
\end{eqnarray}
The term to $n$-th order in~$\hat{P}$ contains all possible electron transfers
generated by $n$~applications of the perturbation~$\hat{T}$;
its contribution is proportional to $(t/U)^n$.
The square-root factor in~(\ref{DEFgamma}) guarantees the size-consistency
of the expansion, i.e., it eliminates the `unconnected' diagrams in
a diagrammatic formulation of the theory\cite{Takahashi}. 
The square root of an operator is understood in terms
of its series expansion, i.e.,
\begin{equation}
\left( \hat{P}_0\hat{P}\hat{P}_0 \right)^{-1/2}
\equiv \hat{P}_0  + \sum_{n=1}^{\infty}
\frac{(2n-1)!!}{(2n)!!}
\left[  
\hat{P}_0 (\hat{P}_0 -\hat{P})\hat{P}_0 
\right]^n \; .
\end{equation}
The Green function for the lower Hubbard band becomes
\begin{equation}
G_{\rm LHB}(\omega ) = \frac{1}{L} \sum_{i,\sigma}
        \left\langle \widetilde{c}_{i,\sigma}^+
            \left[\omega + \hat{h} -\I\eta\right]^{-1} 
         \widetilde{c}_{i,\sigma}\right\rangle \; ,
\label{DefGLHB2ndintermediate}
\end{equation}
where $\langle \ldots \rangle$ now implies the average over
all states~$|\phi_n\rangle$ with no holes and no
double occupancies ($N=L$ electrons),
i.e., the average over all single-spin configurations.

Up to and including second order in $1/U$ a straightforward
expansion gives
\begin{eqnarray}
\hat{\Gamma} &=& \hat{P}_0 
+ \hat{S}^1\hat{T}\hat{P}_0
- \hat{S}^2\hat{T}\hat{P}_0\hat{T}\hat{P}_0
+ \hat{S}^1\hat{T}\hat{S}^1\hat{T}\hat{P}_0 
- \frac{1}{2} \hat{P}_0 \hat{T}\hat{S}^2\hat{T}\hat{P}_0 \; , \\
\widetilde{c}_{i,\sigma} &=& \hat{P}_0  \hat{c}_{i,\sigma} \hat{P}_0 
+ \hat{P}_0  \hat{c}_{i,\sigma} \hat{S}^1\hat{T}\hat{P}_0 
+ \hat{P}_0 \hat{T}\hat{S}^1 \hat{c}_{i,\sigma} \hat{S}^1\hat{T}\hat{P}_0 
+ \hat{P}_0 \hat{c}_{i,\sigma} \hat{S}^1 \hat{T}\hat{S}^1\hat{T}\hat{P}_0 
\nonumber \\
&& -\frac{1}{2} \hat{P}_0 \hat{c}_{i,\sigma} 
             \hat{P}_0 \hat{T}\hat{S}^2\hat{T}\hat{P}_0
- \frac{1}{2} \hat{P}_0 \hat{T}\hat{S}^2\hat{T}\hat{P}_0 
              \hat{c}_{i,\sigma} \hat{P}_0 
\; .
\label{widetildec} 
\end{eqnarray}
In the derivation of~(\ref{widetildec}) we used the fact that 
$\widetilde{c}_{i,\sigma}$ in~(\ref{DefGLHB2ndintermediate}) 
acts on states $|\phi_n\rangle$ with
no holes and no double occupancies.
The transformed Hamiltonian up to second order reads
\begin{equation}
\hat{h} = \frac{U}{4}(L-2\hat{N}) +
\hat{h}_0+ \hat{h}_1 + \hat{h}_2 + \ldots
\label{Hexpandone}
\end{equation}
with
\begin{eqnarray}
\hat{h}_0 &=& \hat{P}_0 \hat{T} \hat{P}_0 \; , \nonumber \\[3pt]
\hat{h}_1 &=& \hat{P}_0 \hat{T} \hat{S}^1 \hat{T} \hat{P}_0 +\frac{L}{2U}\; ,
\label{Hexpandtwo}
\\[3pt]
\hat{h}_2 &=& \hat{P}_0 \hat{T} \hat{S}^1 \hat{T} \hat{S}^1 \hat{T} \hat{P}_0 
-\frac{1}{2} \hat{P}_0 \hat{T} \hat{P}_0 \hat{T} \hat{S}^2 \hat{T} \hat{P}_0 
-\frac{1}{2} \hat{P}_0 \hat{T} \hat{S}^2 \hat{T} \hat{P}_0 \hat{T} \hat{P}_0 
\nonumber \; ,
\end{eqnarray}
where we used $E_0^{(0)}=0$, $E_0^{(1)}=-L/(2U)$, and $E_0^{(2)}=0$ 
for spin-disordered ground states\cite{Takahashi}. 

Including all second-order terms,
the Green function for the lower Hubbard band~(\ref{DefGLHB}) becomes
\begin{equation}
G_2(\omega ) = \frac{1}{L} \sum_{i,\sigma}
        \left\langle \widetilde{c}_{i,\sigma}^+
            \left[\omega +U/2+ \hat{h}_0 + \hat{h}_1 +\hat{h}_2
-\I\eta\right]^{-1} 
         \widetilde{c}_{i,\sigma}\right\rangle \; . \label{DefGLHB2nd}
\end{equation}
For the Falicov--Kimball model, all operators $\hat{T}$ need to be replaced
by $\hat{T}^{\rm FK}$ from~(\ref{FKkineticenergy}), and the spin index 
in~(\ref{widetildec}) and~(\ref{DefGLHB2nd}) must be dropped.

\subsection{Green Function to Lowest Order}
\label{GFlowestORDER}

As a first application we derive the well-known lowest-order result
for the local hole Green function,
\begin{equation}
G_0(z) =     
\frac{1}{L} \sum_{i,\sigma}   \left\langle \hat{c}_{i,\sigma}^+
            \left[z +\hat{h}_0\right]^{-1} 
         \hat{c}_{i,\sigma}\right\rangle \label{Gzerodef} \; ,
\end{equation}
where $z$~is a complex variable which we shall put 
to $z=\omega+U/2-\I\eta$ at the end of the calculation.

A series expansion for $G_0(z)$ in terms of $\hat{h}_0/z$ converges
for $|z|>2$,
\begin{equation}
G_0(z) =     \frac{1}{z} \sum_{n=0}^{\infty} 
\frac{1}{L} \sum_{i,\sigma}   \left\langle \hat{c}_{i,\sigma}^+
            \left(-\frac{\hat{h}_0}{z}\right)^n
         \hat{c}_{i,\sigma}\right\rangle \; .
\end{equation}
The term in $n$-th order describes the creation of a hole at site~$i$
which moves through the lattice in such a way that it returns after
$n$~nearest-neighbor hops. Therefore, $n$~is even, and the spin background is
restored after the hole has returned to~$i$.

Obviously, the hole can return to~$i$ more than once during
its excursions. We define $S(z)$ as the contributions to the series
such that the hole does {\sl not\/} return in intermediate steps.
Then,
\begin{equation}
G_0(z) =  \frac{1}{z} \sum_{n=0}^{\infty} [S(z)]^n = \frac{1}{z(1-S(z))}\; . 
\label{Gzzeroseriesexpansion}
\end{equation}
On a Bethe lattice there are no loops. Therefore, the excursions which
contribute to $S(z)$ start with a jump of the hole
to one of the $Z$~nearest neighbors~$i+\tau$
which serves as a starting point for arbitrary excursions which 
return to~$i+\tau$ and avoid~$i$.
The latter constraint is irrelevant for $Z\to\infty$. The last jump
takes the hole from site~$i+\tau$ back to~$i$. This implies
\begin{equation}
S(z) = Z \left( \frac{1}{z \sqrt{Z}}\right)^2
\sum_{m=0}^{\infty} [S(z)]^m = \frac{1}{z^2(1-S(z))} \; .
\label{Sofz}
\end{equation}
This quadratic equation is readily solved to give
\begin{equation}
G_0(z) = zS(z) = \frac{1}{2} \left[ z + \sqrt{z^2-4}\, \right] \;,
\label{Gzeroresults}
\end{equation}
in agreement with the exact solution\cite{MVSchmit}.
The sign has to be chosen to fulfill the proper limit for 
$\Re\{z\}\to-\infty$.

We perform an analytical continuation from $|z|>2$ to the 
whole complex plane so that the formula~(\ref{Gzeroresults})
is valid for all~$z$.
In particular, for $z=\omega+U/2-\I\eta$ the density of states of
the lower Hubbard band becomes 
\begin{eqnarray}
D_0(\omega) &=& \frac{1}{2\pi} \sqrt{4 - (\omega+U/2)^2\, } 
\quad , \quad |\omega+U/2| \leq 2 \nonumber \; ,\\
&=&\rho(\omega+U/2) \; .
\label{Dzero}
\end{eqnarray}
This result does not come as a surprise because the motion of the hole
restores the spin configuration. Therefore, the local density
of states of the non-interacting system~(\ref{rhozero}) is 
reproduced apart from the energy shift~$U/2$.
The one-particle gap $\Delta$~(\ref{Delta}) becomes
\begin{equation}
\Delta_0(U) = U-4
\label{deltazeroHUB}
\end{equation}
to leading order in~$1/U$ because $\mu_{\rm LHB}^+=-U/2+2$.

For the Falicov--Kimball model we proceed along the same lines.
Note, however, that (i)~the creation
of a hole at site~$i$ has only probability $p=1/2$,
cf.~(\ref{Gzzeroseriesexpansion}), and,~(ii),
the probability for a hole hop from~$i$ to $i+\tau$
is also reduced by a factor of~$p$,
cf.~(\ref{Sofz}). Thus,
\begin{equation}
G^{\rm FK}_0(z) =  \frac{1}{2} 
\frac{1}{z} \sum_{n=0}^{\infty} [S^{\rm FK}(z)]^n = 
\frac{1}{2z(1-S^{\rm FK}(z))}\; , 
\label{GzzeroseriesexpansionFK}
\end{equation}
and
\begin{equation}
S^{\rm FK}(z) = \frac{1}{2} Z \left( \frac{1}{z \sqrt{Z}}\right)^2
\sum_{m=0}^{\infty} [S^{\rm FK} (z)]^m = \frac{1}{2 z^2(1-S^{\rm FK}(z))} \; .
\label{SofzFK}
\end{equation}
Therefore,
\begin{equation}
G^{\rm FK}_0(z) = zS^{\rm FK}(z) = 
\frac{1}{2} \left[ z + \sqrt{z^2-2}\, \right] \;, 
\label{GzeroresultsFK}
\end{equation}
in agreement with the exact solution\cite{PvD}; see Sect.~\ref{Sec:FKexact}

After the analytical continuation to all $z$
and the replacement $z=\omega+U/2-\I\eta$,
we find for the density of states of the lower Hubbard band 
in the Falicov--Kimball model
\begin{eqnarray}
D^{\rm FK}_0(\omega) &=& \rho^{\rm FK}(\omega+U/2)  \nonumber \\
&=& \frac{1}{2\pi} \sqrt{2 - (\omega+U/2)^2\, } 
\quad , \quad |\omega+U/2| \leq \sqrt{2} \; .
\label{DzeroFK} 
\end{eqnarray}
The one-particle gap $\Delta$~(\ref{Delta}) becomes
\begin{equation}
\Delta_0^{\rm FK}(U) = U-2\sqrt{2}
\label{deltazeroFK}
\end{equation}
to leading order in~$1/U$ because $[\mu^{\rm FK}_{\rm LHB}]^+
=-U/2+\sqrt{2}$.

\section{FALICOV--KIMBALL MODEL}
\label{SecIV}

The calculation of the higher-order terms of the Green function
leads to a resummation problem. As its solution
we propose the scheme of an
{\sl effective operator for the hole motion}. The validity of
our approach is checked against the exact solution
of the Falicov--Kimball model\cite{PvD}. We shall drop
the index `FK' in this section.

\subsection{Exact One-Particle Green Function}
\label{Sec:FKexact}

As shown in Ref.~\onlinecite{PvD}, the one-particle Green
function for all $U$ is obtained from the solution
of the cubic equation\cite{PvDwarning}
\begin{equation}
[G(\omega)]^3 -2\omega[G(\omega)]^2 
+G(\omega)(1+\omega^2-U^2/4) -\omega=0 \; .
\label{exactGFFK}
\end{equation}
The gap is explicitly given by\cite{PvD}
\begin{equation}
\Delta(U)= 2\left\{ \frac{2}{U^2} \left[ 
-\frac{1}{4} + \frac{5U^2}{4}+\frac{U^4}{8} 
-\sqrt{\frac{1}{16}+\frac{3U^2}{8}+\frac{3U^4}{4}+\frac{U^6}{2}}\,
\right]\right\}^{1/2}
\label{fullgapFK}
\end{equation}
for $U\geq U^{\rm FK}_{\rm c}=2$.
The $1/U$~expansion of the gap to first and second order reads
\begin{eqnarray}
\Delta_1(U) &=& U-2\sqrt{2}+\frac{1}{U} \; ,  \label{gapFK1storderEX} \\
\Delta_2(U) &=& U-2\sqrt{2}+\frac{1}{U}+\frac{\sqrt{2}}{2}\frac{1}{U^2} \; .
\label{gapFK2ndorder}
\end{eqnarray}
All higher-order corrections turn out to be positive.
It is therefore seen that the Hubbard bands tend to `repel'
each other in the Falicov-Kimball model so that the gap bends
upwards when the interaction is reduced towards the transition.

\begin{figure}[ht]
\vspace*{-0.4cm}
\centerline{\includegraphics[height=12cm,width=10cm,angle=-90]{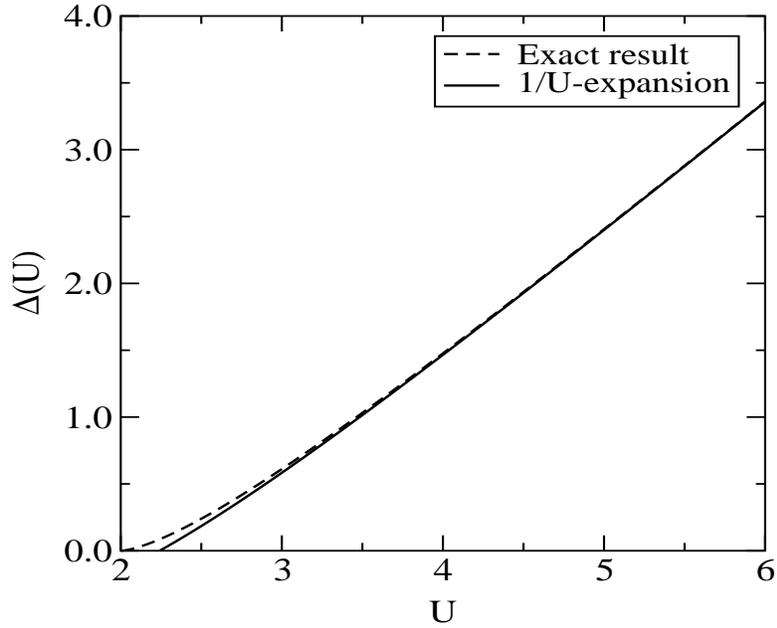}}
\caption{Mott--Hubbard gap for one-particle excitations in the
Falicov--Kimball model. The exact result
(dashed line) and the result from the expansion to second order 
in~$1/U$ (solid line) are shown.}  
\label{Fig:gapFK}
\end{figure}

The large-$U$ expansion can also be used to make predictions
about the value of the critical interaction strength~$U_{\rm c}$
for the Mott--Hubbard transition. From $\Delta_n(U)=0$ ($n=0,1,2$)
in~(\ref{deltazeroFK}),~(\ref{gapFK1storderEX}),~(\ref{gapFK2ndorder})
we find
\begin{equation}
U_{\rm c}^{(0)} = 2.83 \quad ,\quad
U_{\rm c}^{(1)} = 2.41 \quad ,\quad
U_{\rm c}^{(2)} = 2.24 \quad ,\quad
\ldots \quad ,\quad
U_{\rm c} = 2 \; .
\end{equation}
It is seen that the second-order result is only about 10~percent off the
exact result.
In Fig.~\ref{Fig:gapFK} we compare the result to second 
order~(\ref{gapFK2ndorder}) with the exact result~(\ref{fullgapFK}).
The agreement is very good for $U\geq 3=(3/2) U_{\rm c}$.
As we shall show in Sect.~\ref{Sec:DOSFK}
the $1/U$~expansion works equally well for the density of states.

We now set $\omega=z-U/2$ in~(\ref{exactGFFK})
and consider $|z|={\cal O}(1)$ for
the lower Hubbard band. This leads to
\begin{equation}
[G(z)]^3 -2z[G(z)]^2 +G(z)(1+z^2)-z 
+U\left[ [G(z)]^2 -zG(z) +1/2\right] = 0 \; , 
\label{exactGFFKtransformed}
\end{equation}
from which the solution to lowest order in $1/U$ immediately follows
as $G_0(z)$, Eq.~(\ref{GzeroresultsFK}).

\subsection{Green Function to First Order}
\label{1storderFK}

The first-order corrections $G_1(\omega)$ stem from two sources,
(i)~the expansion of the Fermi operators~(\ref{widetildec}),
which are henceforth called {\sl shape-correction terms}, 
and,~(ii), the expansion of the Hamiltonian~(\ref{Hexpandone}).
The latter terms will lead to the gap renormalization problem.

\subsubsection{Shape-correction Terms}

We start from~(\ref{DefGLHB2nd}).
The following first-order term arises due to the expansion~(\ref{widetildec}),
\begin{equation}
G_{1,\alpha}(z) =  
\frac{1}{L} \sum_{i}   \left\langle \hat{T}\hat{S}^1 \hat{c}_{i}^+
    \hat{P}_0 \left[z +\hat{h}_0\right]^{-1} 
         \hat{c}_{i}\right\rangle \label{Gonealphadef} ,
\end{equation}
together with its Hermitian conjugate.
The operator $\hat{c}_{i}$ in~(\ref{Gonealphadef})
can only act if there is no immobile $f$~electron at site~$i$.
On the other hand, the operator $\hat{S}^1 \hat{c}_i^+$ 
requires an immobile $f$~electron at~$i$. Therefore,
the first-order shape-correction terms vanish for the
Falicov--Kimball model.

\subsubsection{Gap Renormalization Problem}

To first order in $1/U$ we are left with the calculation of 
\begin{eqnarray}
G_1(z) 
&=& \frac{1}{L} \sum_{i}   \left\langle \hat{c}_{i}^+
            \left[z +\hat{h}_0 +\hat{h}_1 \right]^{-1} 
                 \hat{c}_{i} \right\rangle 
\label{GFKfirstmain}
\\
&=& G_0(z) + \widetilde{G}_1(z) + {\cal O}\left(U^{-2}\right)\nonumber \; ,\\
\widetilde{G}_1(z) &= & - \frac{1}{L} \sum_{i}   \left\langle \hat{c}_{i}^+
            \left[z +\hat{h}_0\right]^{-1} 
         \hat{h}_1 \left[z +\hat{h}_0\right]^{-1} 
        \hat{c}_{i} \right\rangle  
\label{GFKfirst}
\end{eqnarray}
with $z=\omega+U/2-\I\eta$ at the end of the calculation.

The operator $\hat{h}_1$ in~(\ref{Hexpandtwo}) can be decomposed into 
two-site and three-site contributions\cite{Gebhardbook},
$\hat{h}_1= \hat{h}_1^{\rm 2s} + \hat{h}_1^{\rm 3s}$,
\begin{eqnarray}
\hat{h}_1^{\rm 2s} 
&=&  
- \frac{1}{Z U} \hat{P}_0 \sum_{l,\tau} 
\left(\hat{c}_{l+\tau}^+ \hat{c}_{l+\tau} \hat{f}_{l}^+\hat{f}_{l} 
-\frac{1}{4}\right) \hat{P}_0 
\; , \nonumber \\
\hat{h}_1^{\rm 3s} &=&  - \frac{1}{Z U}  \hat{P}_0
\sum_{l,\tau_1\neq\tau_2} 
\hat{c}_{l+\tau_2}^+ \hat{c}_{l} 
\hat{f}_{l}^+\hat{f}_{l}
\hat{c}_{l}^+ \hat{c}_{l+\tau_1} \hat{P}_0
\; . 
\end{eqnarray}
The three-site terms do not contribute to $\widetilde{G}_1(z)$: The operator
$\hat{h}_1^{\rm 3s}$ allows the hole at $l+\tau_2$ to jump
over the immobile $f$~electron at site~$l$ to reach site $l+\tau_1$.
However, the site $l$ cannot be passed again
by the simple hole motion via $\hat{h}_0$. Therefore, the hole,
created at site~$i$, cannot return.

The two-site terms give rise to the constant $1/(2U)$ when applied to
a state with a single hole. Thus,
\begin{equation}
\widetilde{G}_1(z)  =  - \frac{1}{2U} \frac{1}{L} \sum_{i}   
\left\langle \hat{c}_{i}^+
            \left[z +\hat{h}_0\right]^{-2} 
        \hat{c}_{i} \right\rangle 
= \frac{1}{4U} \left[ 1 + \frac{z}{\sqrt{z^2-2}}\right] \; .
\label{DivergenceGFK}
\end{equation}
It is evident from~(\ref{DivergenceGFK}) that the first-order term
{\sl diverges\/} at $\omega=-U/2+\sqrt{2}=[\mu_{\rm LHB}]_0^+$, where
the density of states to lowest order~(\ref{DzeroFK}) becomes zero.
This had to be expected because the density of states
and the gap are indeed renormalized
to first order in $1/U$, see~(\ref{gapFK1storderEX}).
As is well known from standard perturbation theory in~$U$,
a shift of the one-particle resonances cannot be 
generated by the calculation of finite-order corrections
to the Green function but requires an appropriate summation
of an infinite series as performed, e.g., via the Dyson equation\cite{Fetter}.
For our case of a degenerate ground state, there are no standard 
means of carrying out such a resummation. 
In the following paragraph we present the solution for the insulating phase
of the Falicov--Kimball model on the $Z\to\infty$ Bethe lattice 
in terms of an effective operator for the hole motion.

\subsubsection{Effective Operator for the Hole Motion}

For a Mott--Hubbard insulator 
the motion of the hole through the system does not
alter the positions of the immobile electrons. Moreover, the
background does not have dynamics of its own. To lowest order,
the hole dynamics is governed by~$\hat{h}_0$, and we expect that
higher orders in the $1/U$~expansion correspond to a modified
hole hopping. These modifications can be expressed
in terms of an effective operator for the hole motion.
In fact, by comparing~(\ref{GFKfirst}) and~(\ref{DivergenceGFK}),
the effective hopping to first order must take the form
\begin{equation}
\hat{h}_1 \to \hat{h}_1^{\rm eff} =\frac{1}{2U} 
\label{heff1FK} \; .
\end{equation}
We insert this result into~(\ref{GFKfirstmain}) and find 
\begin{equation}
G_1(z) = G_0(z+1/(2U)) \; .
\end{equation}
The first-order contribution is merely a band shift by~$1/(2U)$.

\subsubsection{Density of States}

The density of states for the lower Hubbard band
is simply shifted downwards in energy by $1/(2U)$,
\begin{equation}
D_1(\omega) = \frac{1}{2\pi} \sqrt{2-
\left(\omega+\frac{U}{2}+\frac{1}{2U}\right)^2} 
\quad , \quad \left|\omega+\frac{U}{2}+\frac{1}{2U}\right| 
\leq \sqrt{2} \; .
\end{equation}
Therefore, the gap to first order becomes
\begin{equation}
\Delta_1(U)=U-2\sqrt{2}+\frac{1}{U} \; , 
\label{GapFK1storder}
\end{equation}
because $[\mu_{\rm LHB}]_1^+=-U/2+\sqrt{2}-1/(2U)$,
up to and including first order in~$1/U$.
Eq.~(\ref{GapFK1storder}) agrees 
with the exact result~(\ref{gapFK1storderEX}).

\subsection{Green Function to Second Order}

\subsubsection{Shape-correction Terms}

We again start from~(\ref{DefGLHB2nd}).
The following second-order terms arise due to the expansion~(\ref{widetildec}),
\begin{eqnarray}
G_{2,\alpha}(z) &=&  -
\frac{1}{L} \sum_{i}   \left\langle \hat{T}\hat{S}^1 \hat{c}_{i}^+
    \hat{P}_0 \left[z +\hat{h}_0\right]^{-1} 
\hat{h}_1
 \left[z +\hat{h}_0\right]^{-1} 
         \hat{c}_{i}\right\rangle \label{Gtwoalphadef} \; ,\\
G_{2,\beta}(z) &=&  \frac{1}{L} \sum_{i}   
\left\langle \hat{c}_{i}^+     \hat{P}_0 
\left[z +\hat{h}_0\right]^{-1} 
\hat{P}_0 \hat{c}_{i} 
\hat{S}^1 \hat{T}\hat{S}^1 \hat{T}
\right\rangle \label{Gtwobetadef} \; ,\\
G_{2,\gamma}(z) &=&  \frac{1}{L} \sum_{i}   
\left\langle \hat{c}_{i}^+ 
\hat{P}_0 \left[z +\hat{h}_0\right]^{-1} 
\Bigl[ \hat{P}_0 \hat{T} \hat{S}^1 \hat{c}_{i} \hat{S}^1 \hat{T}
\right. \nonumber \\
&& \hphantom{\frac{1}{L} \sum_{i}   
\biggl\langle  }
\left. -\frac{1}{2} \hat{P}_0 \hat{c}_{i} 
             \hat{P}_0 \hat{T}\hat{S}^2\hat{T}\hat{P}_0
- \frac{1}{2} \hat{P}_0 \hat{T}\hat{S}^2\hat{T}\hat{P}_0 
              \hat{c}_{i} \hat{P}_0 \Bigr]
\right\rangle \label{Gtwogammadef} \; ,\\
G_{2,\delta}(z) &=&  \frac{1}{L} \sum_{i}   
\left\langle \hat{T} \hat{S}^1 \hat{c}_{i}^+ \hat{P}_0 
\left[z +\hat{h}_0\right]^{-1} 
 \hat{P}_0 \hat{c}_{i} \hat{S}^1 \hat{T} 
\right\rangle \label{Gtwodeltadef} \;,
\end{eqnarray}
together with their Hermitian conjugates.
With the help of the same arguments used in Sect.~\ref{1storderFK},
it is not difficult to show that
$G_{2,\alpha}(z) =0$ and $G_{2,\beta}(z) =0$.
Moreover, the series expansion as exemplified in Sect.~\ref{GFlowestORDER}
leads to $G_{2,\gamma}(z) = -G_{2,\delta}(z)$ so that 
the shape-correction terms vanish altogether for the
Falicov--Kimball model also in second order in~$1/U$.

\subsubsection{Effective Operator for the Hole Motion}

To second order in $1/U$ we are left with the calculation of 
\begin{eqnarray}
G_2(z) 
&=& \frac{1}{L} \sum_{i}   \left\langle \hat{c}_{i}^+
            \left[z +\hat{h}_0 +\hat{h}_1 +\hat{h}_2 \right]^{-1} 
                 \hat{c}_{i} \right\rangle 
\label{GFKsecondmain}
\\
&=& G_0(z) + \widetilde{G}_1(z) + \widetilde{G}_2(z) + 
{\cal O}\left(U^{-3}\right) \; , \nonumber 
\end{eqnarray}
which requires the evaluation of
\begin{eqnarray}
\widetilde{G}_2(z) &=&   \frac{1}{L} \sum_{i}   \left\langle \hat{c}_{i}^+
\left[z +\hat{h}_0\right]^{-1} \left[  (\hat{h}_1^{\rm 2s} +
\hat{h}_1^{\rm 3s}) 
 \left[z +\hat{h}_0\right]^{-1}  (\hat{h}_1^{\rm 2s}+ \hat{h}_1^{\rm 3s})   
- \hat{h}_2 \right] \right. \nonumber \\
&& \hphantom{\frac{1}{L} \sum_{i} \biggl\langle}
\left. \left[z +\hat{h}_0\right]^{-1} 
        \hat{c}_{i} \right\rangle  
\label{GFKsecond}
\end{eqnarray}
with $z=\omega+U/2-\I\eta$ at the end of the calculation.

With the help of the series expansion, see Sect.~\ref{GFlowestORDER},
the result for $\widetilde{G}_2(z)$ can be cast into the form
\begin{eqnarray}
U^2\widetilde{G}_2(z) & =& \frac{1}{4}
 \frac{1}{L} \sum_{i}   \left\langle \hat{c}_{i}^+
\left[z +\hat{h}_0\right]^{-3} 
        \hat{c}_{i} \right\rangle 
%\nonumber \\
%
% && 
+ \frac{G_0(z)}{2}
 \frac{1}{L} \sum_{i}   \left\langle \hat{c}_{i}^+
\left[z +\hat{h}_0\right]^{-2} 
        \hat{c}_{i} \right\rangle \nonumber \\
&& - G_0(z)
 \frac{1}{L} \sum_{i}   \left\langle \hat{c}_{i}^+
\left[z +\hat{h}_0\right]^{-2} 
        \hat{c}_{i} \right\rangle \label{FK2ndorderGtilde} \; .
\end{eqnarray}
The first term arises from the combination of the two factors
$\hat{h}_1^{\rm 2s}$, the second term stems from
the two factors~$\hat{h}_1^{\rm 3s}$ (mixed terms are zero),
and the third term comes from~$\hat{h}_2$ in~(\ref{GFKsecond}).
We note that
\begin{equation}
 \frac{1}{L} \sum_{i}   \left\langle \hat{c}_{i}^+
\hat{h}_0 \left[z +\hat{h}_0\right]^{-2} 
        \hat{c}_{i} \right\rangle 
= -2 G_0(z)  \frac{1}{L} \sum_{i}   \left\langle \hat{c}_{i}^+
 \left[z +\hat{h}_0\right]^{-2} \hat{c}_{i} \right\rangle \;,
\end{equation}
so that the replacements~(\ref{heff1FK}) and
\begin{equation}
\hat{h}_2 \to \hat{h}_2^{\rm eff} =-\frac{1}{4U^2} \hat{h}_0
\label{heff2FK}
\end{equation}
in~(\ref{GFKsecondmain}) reproduce~(\ref{FK2ndorderGtilde}).
Up to and including second order we may therefore write
\begin{equation}
G_2(z) =  \frac{1}{L} \sum_{i}   \left\langle \hat{c}_{i}^+
            \left[z +\hat{h}_0 + 1/(2U) - \hat{h}_0/(4U^2)
 \right]^{-1} 
                 \hat{c}_{i} \right\rangle \; .
\end{equation}
The second-order corrections result in a shift and a 
narrowing of the Hubbard bands.

\begin{figure}[ht]
\vspace*{-0.3cm}
\centerline{\includegraphics[height=11cm,width=10cm,angle=-90]{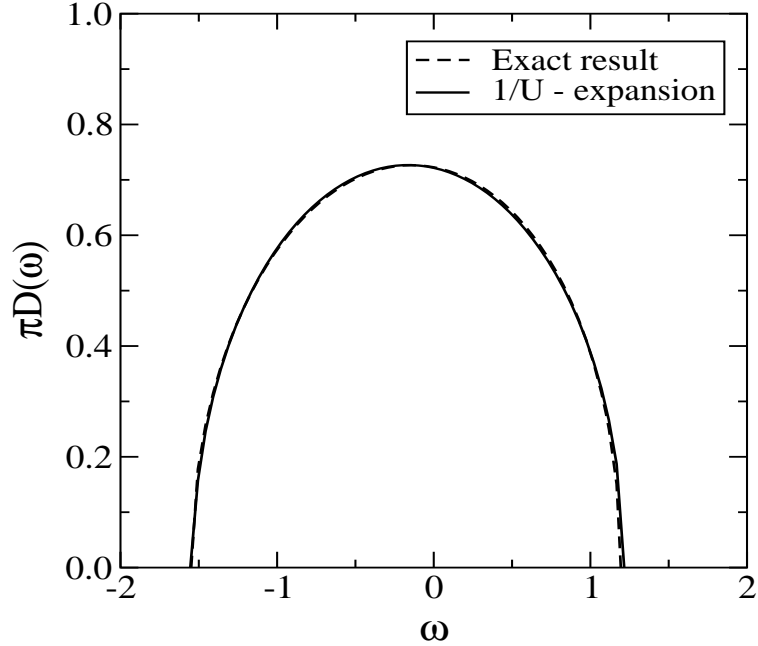}}
\caption{Density of states of the lower Hubbard band in the
Falicov--Kimball model, shifted by $U/2$, for $U=3=1.5 U_{\rm c}$. 
The exact result
(dashed line) and the result from the expansion to second order 
in~$1/U$ (solid line) are shown.}  
\label{Fig:DOSFK}
\end{figure}

\subsubsection{Density of States}
\label{Sec:DOSFK}

Since we know the local density of states of~$\hat{h}_0$, eq.~(\ref{rhoFKdef}),
the density of states of the lower Hubbard band becomes
\begin{eqnarray}
D_2(\omega) & = & \int_{-\infty}^{\infty} d\epsilon 
\rho(\epsilon)\delta\left(\omega+\frac{U}{2}+
\epsilon+\frac{1}{2U}-\frac{\epsilon}{4U^2}\right)
\nonumber \\[3pt]
&=& \frac{1}{1-1/(4U^2)} \rho\left(\frac{\omega+U/2+1/(2U)}{1-1/(4U^2)}\right)
\label{FinalDFK}\; , 
\end{eqnarray}
which explicitly reads
\begin{eqnarray}
D_2(\omega) &=& \frac{1}{2\pi} \left(1-\frac{1}{4U^2}\right)^{-2}  
\sqrt{2\left(1-\frac{1}{4U^2}\right)^2  - 
\left( \omega +\frac{U}{2} +\frac{1}{2U} \right)^2}\; , \nonumber \\[3pt]
&& \hfill 
\left| \omega +\frac{U}{2} +\frac{1}{2U} \right| \leq 
\sqrt{2} -\frac{\sqrt{2}}{4U^2} \label{FinalDFKsecond}
\end{eqnarray}
up to and including second order in~$1/U$.
Therefore, the gap to second order becomes
\begin{equation}
\Delta_2(U)=U-2\sqrt{2}+\frac{1}{U}+\frac{\sqrt{2}}{2U^2} \; , 
\label{GapFKsecondorderexpansion}
\end{equation}
because $[\mu^{\rm FK}_{\rm LHB}]_2^+=
-U/2+\sqrt{2}-1/(2U)-\sqrt{2}/(4U^2)$
up to and including second order in~$1/U$.
Eq.~(\ref{GapFKsecondorderexpansion}) agrees 
with the exact result~(\ref{gapFK2ndorder}).

As seen from Fig.~\ref{Fig:DOSFK}, there is hardly any noticeable
difference between the exact density of states 
from~(\ref{DefGLHB}),~(\ref{exactGFFK}),
and the second-order
result~(\ref{FinalDFKsecond}). 
The favorable comparison shows
that our~$1/U$ expansion works indeed, and that 
is reliable down to $U_{\rm c}/U\approx 2/3$ in the Falicov--Kimball model.
We are confident that the convergence will be similarly good
for the Hubbard model.

\section{HUBBARD MODEL}
\label{SecV}

In Sect.~\ref{SecIV} we have shown that,
as far as the Green function for the lower Hubbard band is concerned,
the operator $\hat{h}$ in~(\ref{deflittleh}) can be replaced by
an effective operator for the hole motion, $\hat{h}^{\rm eff}$, which
can be written in terms of a polynomial in~$\hat{h}_0$.
This was possible because the $f$~electrons are immobile and
randomly distributed over the lattice in the Mott--Hubbard 
insulating phase so that their are no spatial correlations between
the mobile and immobile electrons.

Qualitatively the same applies to the Mott--Hubbard insulator
in the Hubbard model: the spin degrees of freedom are not dynamic
and spatial correlations between the electron spins are absent.
Therefore, the replacement of the operator~$\hat{h}$ by
an effective operator for the hole motion, $\hat{h}^{\rm eff}$,
is permissible as we have verified from the explicit series expansion as
used in Sect.~\ref{GFlowestORDER}

\subsection{Green Function to First Order}
\label{1storderHubbard}

\subsubsection{Shape-correction Terms}

We start from~(\ref{DefGLHB2nd}).
The following first-order term arises due to the expansion~(\ref{widetildec}),
\begin{equation}
G_{1,\alpha}(z) =
\frac{1}{L} \sum_{i,\sigma}   \left\langle \hat{T}\hat{S}^1 
\hat{c}_{i,\sigma}^+
    \hat{P}_0 \left[z +\hat{h}_0\right]^{-1} 
         \hat{c}_{i,\sigma}\right\rangle \label{GonealphadefHUB} \; ,
\end{equation}
together with its Hermitian conjugate. 
The operator $\hat{T}\hat{S}^1 $ involves one additional hole transfer
so that necessarily has the form
\begin{equation}
G_{1,\alpha}(z) = \frac{1}{L} \sum_{i,\sigma}   \left\langle 
\hat{c}_{i,\sigma}^+ (s_{1} \hat{h}_0) 
    \hat{P}_0 \left[z +\hat{h}_0\right]^{-1} 
         \hat{c}_{i,\sigma}\right\rangle 
\label{Gonealphasimplified} \; .
\end{equation}
Now that we only need to determine
a single real parameter~$s_1$, we expand the right-hand-sides
of~(\ref{GonealphadefHUB}) and~(\ref{Gonealphasimplified}) in $1/z$
and compare the terms to order $1/z^2$:
\begin{equation}
\frac{1}{U} \frac{1}{L} \sum_{i,\sigma}   \left\langle 
\hat{T}\hat{D} \hat{c}_{i,\sigma}^+ \hat{h}_0
         \hat{c}_{i,\sigma}\right\rangle 
= - s_1 
\frac{1}{L} \sum_{i,\sigma}   \left\langle
 \hat{c}_{i,\sigma}^+ (\hat{h}_0)^2
         \hat{c}_{i,\sigma}\right\rangle \; ,
\end{equation}
which reduces to
\begin{equation}
\frac{1}{2U} =-s_1\; .
\end{equation}
Therefore, the operator for the shape-correction term
to first order is given by
\begin{equation}
\widetilde{c}_{i,\sigma} = \hat{P}_0  
\left[ 1 -\frac{1}{2U} \hat{h}_0 \right] \hat{c}_{i,\sigma} \hat{P}_0 
+{\cal O}\left(U^{-2}\right) \; .
\label{shapeHubbard1st}
\end{equation}

\subsubsection{Effective Operator for the Hole Motion}

To first order in $1/U$ we are left with the calculation of 
\begin{eqnarray}
G_{1,\beta}(z) 
&=& \frac{1}{L} \sum_{i,\sigma}   \left\langle \hat{c}_{i,\sigma}^+
            \left[z +\hat{h}_0 +\hat{h}_1 \right]^{-1} 
                 \hat{c}_{i,\sigma} \right\rangle 
\label{Gfirstmain}
\\
&=& G_0(z) + \widetilde{G}_1(z) + {\cal O}\left(U^{-2}\right) \; ,
\nonumber
\end{eqnarray}
which requires the evaluation of 
\begin{equation}
\widetilde{G}_1(z) =  - \frac{1}{L} \sum_{i,\sigma}   
\left\langle \hat{c}_{i,\sigma}^+
            \left[z +\hat{h}_0\right]^{-1} 
         \hat{h}_1 \left[z +\hat{h}_0\right]^{-1} 
        \hat{c}_{i,\sigma} \right\rangle  
\label{Gfirst}
\end{equation}
with $z=\omega+U/2-\I\eta$ at the end of the calculation.

As usual, the operator $\hat{h}_1$ 
in~(\ref{Hexpandtwo}) can be decomposed into 
two-site and three-site contributions\cite{Gebhardbook},
$\hat{h}_1 = \hat{h}_1^{\rm 2s} + \hat{h}_1^{\rm 3s}$,
\begin{eqnarray}
\hat{h}_1^{\rm 2s} &=& \frac{1}{Z} \hat{P}_0  \sum_{l,\tau} 
\frac{2}{U}\left[ \hat{\vec{S}}_{l}\hat{\vec{S}}_{l+\tau}
-\frac{1}{4} \left(\hat{n}_{l}\hat{n}_{l+\tau}-1\right)\right]  
\hat{P}_0 \; , \nonumber \\
\hat{h}_1^{\rm 3s} &=&  - \frac{1}{Z U}  \hat{P}_0
\sum_{l,\tau_1\neq\tau_2;\sigma,\sigma'} \lambda_{\sigma\sigma'}
\hat{c}_{l+\tau_2,\sigma}^+ \hat{c}_{l,-\sigma}^+
\hat{c}_{l,-\sigma'} \hat{c}_{l+\tau_1,\sigma'} \hat{P}_0
\; ,
\end{eqnarray}
where $\lambda_{\sigma\sigma'}=+1 (-1)$ for $\sigma=\sigma'$ 
($\sigma=-\sigma'$).
The operator $\hat{h}_1^{\rm 2s}$ does not move the hole whereas
the operator $\hat{h}_1^{\rm 3s}$ moves the hole by two sites.
Therefore, the replacement
\begin{equation}
\widetilde{G}_{1}(z) = 
- \frac{1}{L} \sum_{i,\sigma} \left\langle \hat{c}_{i,\sigma}^+ 
\left(g_{1;2} (\hat{h}_0)^2 + g_{1;0}\right) 
    \hat{P}_0 \left[z +\hat{h}_0\right]^{-2} 
         \hat{c}_{i,\sigma}\right\rangle 
\label{Gonetildesimplified}
\end{equation}
will lead to the desired result.
The real parameters~$g_{1;2}$ and $g_{1;0}$ are determined as
before. We expand the right-hand-sides
of~(\ref{Gfirst}) and~(\ref{Gonetildesimplified}) in $1/z$
and compare the terms to order $1/z^2$,
\begin{equation}
\frac{1}{L} \sum_{i,\sigma} \left\langle \hat{c}_{i,\sigma}^+ 
\hat{h}_1
        \hat{c}_{i,\sigma} \right\rangle  
=
\frac{1}{L} \sum_{i,\sigma} \left\langle \hat{c}_{i,\sigma}^+ 
\left(g_{1;2} (\hat{h}_0)^2 + g_{1;0}\right) 
        \hat{c}_{i,\sigma} \right\rangle  \; ,
\end{equation}
and to order $1/z^4$,
{\arraycolsep=0pt\begin{eqnarray}
\frac{1}{L} \sum_{i,\sigma}  \left\langle \hat{c}_{i,\sigma}^+
\Bigl( (\hat{h}_0)^2\hat{h}_1+ \right. &&\left. \hat{h}_1 (\hat{h}_0)^2 
+ \hat{h}_0\hat{h}_1\hat{h}_0\Bigr)
        \hat{c}_{i,\sigma} \right\rangle 
= \nonumber \\
&& 3 \frac{1}{L} \sum_{i,\sigma} \left\langle \hat{c}_{i,\sigma}^+ 
\left(g_{1;2} (\hat{h}_0)^4 + g_{1;0}(\hat{h}_0)^2\right) 
        \hat{c}_{i,\sigma} \right\rangle  \; .
\end{eqnarray}}%
The terms are readily calculated to give
\begin{eqnarray}
\frac{1}{U} &=& g_{1;2} + g_{1;0} \; ,\\[3pt]
\frac{3}{2U} &=& 3(2 g_{1;2} + g_{1;0}) \; ,
\end{eqnarray}
where we repeatedly made use of 
the fact that there are no spin correlations in the
Mott--Hubbard insulator.
The solution of this set of equations gives
$g_{1;2}=-1/(2U)$ and $g_{1;0}=3/(2U)$ so that the replacement
\begin{equation}
\hat{h}_1 \to \hat{h}_1^{\rm eff} =-\frac{1}{2U} (\hat{h}_0)^2 + \frac{3}{2U} 
\label{honeeffHUB}
\end{equation}
is valid in~(\ref{Gfirst}).
Combining~(\ref{honeeffHUB}) into~(\ref{Gfirstmain}) 
and~(\ref{shapeHubbard1st}),
the Green function up to and including first order can be written as 
\begin{equation}
G_1(z) =  \frac{1}{L} \sum_{i,\sigma}   \left\langle \hat{c}_{i,\sigma}^+
           \left(1-\frac{\hat{h}_0}{U}\right)
 \left[z +\hat{h}_0 + \frac{3- (\hat{h}_0)^2}{2U}
 \right]^{-1} 
\hat{c}_{i,\sigma} \right\rangle \; .
\label{GFHubbfirstfinal}
\end{equation}

\subsubsection{Density of States}
\label{Sec:DOS}

Since we know the local density of states of~$\hat{h}_0$, eq.~(\ref{Dzero}),
the density of states of the lower Hubbard band becomes
\begin{eqnarray}
D_1(\omega) & = & \int_{-\infty}^{\infty} d\epsilon 
\rho(\epsilon)\left(1-\frac{\epsilon}{U}\right)
\delta\left(\omega+\frac{U}{2}+\epsilon+\frac{3-\epsilon^2}{2U}\right)
\nonumber \\[3pt]
&=& \rho\left(U-\sqrt{2U^2+2U\omega +3}\right) 
\label{FinalDone}\\[3pt]
&=& \frac{1}{2\pi} \left[ 4 - \left(U-\sqrt{2 U^2+2U\omega +3}\right)^2
\right]^{1/2}
\; , \; \left|\omega+\frac{U}{2}-\frac{1}{2U}\right| \leq 2
\; ,\nonumber 
\end{eqnarray}
up to and including first order in~$1/U$.
Therefore, the gap to first order becomes
\begin{equation}
\Delta_1(U)=U-4-\frac{1}{U} \; , 
\label{Gap1storder}
\end{equation}
because $\mu_{\rm LHB}^+=-U/2+2+1/(2U)$,
up to and including first order in~$1/U$.

\subsection{Green Function to Second Order}

\subsubsection{Shape-correction Terms}

We again start from~(\ref{DefGLHB2nd}).
The following second-order terms arise due to the expansion~(\ref{widetildec}),
\begin{eqnarray}
G_{2,\alpha}(z) &=& 
- \frac{1}{L} \sum_{i,\sigma}   \left\langle \hat{T}\hat{S}^1 
\hat{c}_{i,\sigma}^+
    \hat{P}_0 \left[z +\hat{h}_0\right]^{-1} 
\hat{h}_1
 \left[z +\hat{h}_0\right]^{-1} 
         \hat{c}_{i,\sigma}\right\rangle 
\nonumber \\
&=& 
- \frac{1}{L} \sum_{i,\sigma}   \left\langle 
\hat{c}_{i,\sigma}^+ 
\left( s_{2;3}(\hat{h}_0)^3 + s_{2;1}\hat{h}_0\right)
\left[z +\hat{h}_0\right]^{-2} 
         \hat{c}_{i,\sigma}\right\rangle \label{GtwoalphadefHUB} 
\; ,\\
G_{2,\beta}(z) 
&=&  \frac{1}{L} \sum_{i,\sigma}   
\left\langle \hat{c}_{i,\sigma}^+     \hat{P}_0 
\left[z +\hat{h}_0\right]^{-1} 
\hat{P}_0 \hat{c}_{i,\sigma} 
\hat{S}^1 \hat{T}\hat{S}^1 \hat{T}
\right\rangle \nonumber \\
&=&  \frac{1}{L} \sum_{i,\sigma}   
\left\langle \hat{c}_{i,\sigma}^+   
\left( s_{2;2,\beta}(\hat{h}_0)^2 + s_{2;0,\beta}\right)   
\left[z +\hat{h}_0\right]^{-1} 
\hat{c}_{i,\sigma} 
\right\rangle
\label{GtwobetadefHUB} \; ,\\
G_{2,\gamma}(z) &=&  \frac{1}{L} \sum_{i,\sigma}   
\left\langle \hat{c}_{i,\sigma}^+ 
\hat{P}_0 \left[z +\hat{h}_0\right]^{-1} 
\Bigl[ \hat{P}_0 \hat{T} \hat{S}^1 \hat{c}_{i,\sigma} \hat{S}^1 \hat{T}
\right. \nonumber \\
&& \hphantom{\frac{1}{L} \sum_{i,\sigma}   
\biggl\langle  }
\left. -\frac{1}{2} \hat{P}_0 \hat{c}_{i,\sigma} 
             \hat{P}_0 \hat{T}\hat{S}^2\hat{T}\hat{P}_0
- \frac{1}{2} \hat{P}_0 \hat{T}\hat{S}^2\hat{T}\hat{P}_0 
              \hat{c}_{i,\sigma} \hat{P}_0 \Bigr]
\right\rangle \nonumber \\
&=&  \frac{1}{L} \sum_{i,\sigma}   
\left\langle \hat{c}_{i,\sigma}^+   
\left( s_{2;2,\gamma}(\hat{h}_0)^2 + s_{2;0,\gamma}\right)   
\left[z +\hat{h}_0\right]^{-1} 
\hat{c}_{i,\sigma} 
\right\rangle
\label{GtwogammadefHUB} \; ,\\
G_{2,\delta}(z) &=&  \frac{1}{L} \sum_{i,\sigma}   
\left\langle \hat{T} \hat{S}^1 \hat{c}_{i,\sigma}^+ \hat{P}_0 
\left[z +\hat{h}_0\right]^{-1} 
 \hat{P}_0 \hat{c}_{i,\sigma} \hat{S}^1 \hat{T} 
\right\rangle \nonumber \\
&=&  \frac{1}{L} \sum_{i,\sigma}   
\left\langle \hat{c}_{i,\sigma}^+   
\left( s_{2;2,\delta}(\hat{h}_0)^2 + s_{2;0,\delta}\right)   
\left[z +\hat{h}_0\right]^{-1} 
\hat{c}_{i,\sigma} 
\right\rangle
\label{GtwodeltadefHUB} \;,
\end{eqnarray}
together with the Hermitian conjugates
in~(\ref{GtwoalphadefHUB})--(\ref{GtwogammadefHUB}). As in the previous
section, an expansion in~$1/z$ and a comparison of the two leading
orders fixes the unknown real parameters~$s_{2;r,\nu}$. 
{}From~(\ref{GtwoalphadefHUB})--(\ref{GtwodeltadefHUB}) it follows that
\begin{eqnarray}
-\frac{1}{2} \frac{1}{U^2} = 2(2s_{2;3} + s_{2;1})  \quad &,& 
  - \frac{1}{U^2} = 4(5s_{2;3} + 2s_{2;1})  \; , \\
0 = s_{2;2,\beta} + s_{2;0,\beta}  \quad &,& 
 \frac{1}{4} \frac{1}{U^2} = 2s_{2;2;\beta} + s_{2;0,\beta}  \; , \\
-\frac{1}{2} \frac{1}{U^2} = s_{2;2,\gamma} + s_{2;0,\gamma}  \quad &,& 
-\frac{1}{4}  \frac{1}{U^2} = 2s_{2;2;\gamma} + s_{2;0,\gamma}  \; , \\
\frac{1}{U^2} = s_{2;2,\delta} + s_{2;0,\delta}  \quad &,& 
\frac{5}{4}  \frac{1}{U^2} = 2s_{2;2;\delta} + s_{2;0,\delta}  \; .
\end{eqnarray}
Therefore, $U^2 s_{2;3}=1/4$ and $U^2 s_{2;1}=-3/4$ which is also obtained
if we use~(\ref{shapeHubbard1st}) and~(\ref{honeeffHUB}) 
directly in~(\ref{GtwoalphadefHUB}) so that the first-order shape 
correction term remains unaltered.
Moreover, $U^2 s_{2;2,\beta}=1/4$, $U^2 s_{2;0,\beta}=-1/4$;
$U^2 s_{2;2,\gamma}=1/4$, $U^2 s_{2;0,\gamma}=-3/4$; and
$U^2 s_{2;2,\delta}=1/4$, $U^2 s_{2;0,\delta}=3/4$.
Including the contribution from the Hermitian conjugates, we
find $s_{2,2}=2(s_{2;2,\beta}+s_{2;2,\gamma})+ s_{2;2,\delta}=5/(4U^2)$
and $s_{2,0}=2(s_{2;0,\beta}+s_{2;0,\gamma}) + s_{2;0,\delta}=-5/(4U^2)$.
Therefore,
\begin{equation}
\widetilde{c}_{i,\sigma}^+\widetilde{c}_{i,\sigma}
= \hat{c}_{i,\sigma}^+ \hat{P}_0 
\left[ 
1 -\frac{1}{U} \hat{h}_0 +\frac{5}{4U^2} \left((\hat{h}_0)^2 -1\right)\right]
\hat{c}_{i,\sigma} \hat{P}_0  +{\cal O}\left(U^{-3}\right) \; ,
\label{usefulshape}
\end{equation}
so that 
\begin{equation}
\widetilde{c}_{i,\sigma} = \hat{P}_0  
\left[ 1 -\frac{1}{2U} \hat{h}_0 +\frac{1}{2U^2} (\hat{h}_0)^2 -\frac{5}{8U^2}
\right] \hat{c}_{i,\sigma} \hat{P}_0 
+{\cal O}\left(U^{-3}\right)
\label{shapeHubbard2nd}
\end{equation}
is the shape-correction term up to and including second order in~$1/U$.

\subsubsection{Effective Operator for the Hole Motion}

To second order in $1/U$ we are left with the calculation of 
\begin{eqnarray}
G_{2,\epsilon}(z) 
&=& \frac{1}{L} \sum_{i,\sigma}   \left\langle \hat{c}_{i,\sigma}^+
            \left[z +\hat{h}_0 +\hat{h}_1 +\hat{h}_2 \right]^{-1} 
                 \hat{c}_{i,\sigma} \right\rangle 
\label{GsecondmainHUB}
\\
&=& G_0(z) + \widetilde{G}_1(z) + \widetilde{G}_{2}(z) + 
{\cal O}\left(U^{-3}\right)\nonumber \; ,\\
\widetilde{G}_2(z) &= &  
\frac{1}{L} \sum_{i,\sigma}   
\left\langle \hat{c}_{i,\sigma}^+
\left[z +\hat{h}_0\right]^{-1} \left[ \hat{h}_1 
 \left[z +\hat{h}_0\right]^{-1}  \hat{h}_1   
- \hat{h}_2 \right] \right. \nonumber \\
&& \hphantom{\frac{1}{L} \sum_{i} \biggl\langle}
\left. \left[z +\hat{h}_0\right]^{-1} 
        \hat{c}_{i,\sigma} \right\rangle  
\label{GsecondHUB} \; .
\end{eqnarray}
The contribution to $\widetilde{G}_2(z)$ which involves
two factors $\hat{h}_1$ can be obtained by the replacement
$\hat{h}_1 \to \hat{h}_1^{\rm eff}$~(\ref{honeeffHUB})
so that this term does not give a contribution to~$\hat{h}_2^{\rm eff}$.
For the second contribution we write 
{\arraycolsep=0pt\begin{eqnarray}
\frac{1}{L} \sum_{i,\sigma}   
\left\langle \hat{c}_{i,\sigma}^+
\left[z +\hat{h}_0\right]^{-1} \right. && \left. \hat{h}_2 
 \left[z +\hat{h}_0\right]^{-1} 
        \hat{c}_{i,\sigma} \right\rangle  
= \\
&& \frac{1}{L} \sum_{i,\sigma}   
\left\langle \hat{c}_{i,\sigma}^+ 
\left( g_{2;3}(\hat{h}_0)^3 + g_{2;1}\hat{h}_0 \right)
 \left[z +\hat{h}_0\right]^{-2} 
        \hat{c}_{i,\sigma} \right\rangle  \; .  \nonumber 
\end{eqnarray}}%
The equations to ${\cal O}(1/z^3)$ and ${\cal O}(1/z^5)$
result in
\begin{equation}
-\frac{3}{2}\frac{1}{U^2} = 2g_{2;3} + g_{2;1} \quad , \quad 
-\frac{9}{4}\frac{1}{U^2} = 5g_{2;3} + 2g_{2;1} \; ,
\end{equation}
so that $g_{2;3}=3/(4U^2)$, $g_{2;1}=-3/U^2$, and
\begin{equation}
\hat{h}_2 \to \hat{h}_2^{\rm eff} =\frac{3}{4U^2} \hat{h}_0 
\left( (\hat{h}_0)^2 -4\right)
\label{htwoeffHUB}
\end{equation}
is valid in~(\ref{GsecondHUB}).
Combining~(\ref{htwoeffHUB}) into~(\ref{GsecondmainHUB}) 
and~(\ref{usefulshape}),
the Green function up to and including second order can be written as 
\begin{eqnarray}
G_2(z) &=&  \frac{1}{L} \sum_{i,\sigma}   \biggl\langle 
\hat{c}_{i,\sigma}^+
           \left(1-\frac{\hat{h}_0}{U}
+\frac{5[ (\hat{h}_0)^2-1]}{4U^2} \right) \label{GFHubbsecondfinal} \\
&& \hphantom{\frac{1}{L} \sum_{i,\sigma}   \biggl\langle  }
 \left[z +\hat{h}_0 + \frac{3 - (\hat{h}_0)^2}{2U}+
\frac{3\hat{h}_0}{4U^2}\left( (\hat{h}_0)^2-4\right)
 \right]^{-1} 
\hat{c}_{i,\sigma} \biggr\rangle \; .
\nonumber
\end{eqnarray}

\subsubsection{Density of States}

Up to and including second order in~$1/U$
the density of states of the lower Hubbard band becomes
\begin{eqnarray}
D_2(\omega) & = & \int_{-\infty}^{\infty} d\epsilon 
\rho(\epsilon)\left(1-\frac{\epsilon}{U}+
\frac{5(\epsilon^2-1)}{4U^2}\right) \nonumber \\
&& \hphantom{\int_{-\infty}^{\infty} d\epsilon }
\delta\left(\omega+\frac{U}{2}+\epsilon+\frac{3-\epsilon^2}{2U}
+\frac{3\epsilon(\epsilon^2-4)}{4U^2}\right) \label{DOSHUB2ndorder} \\
&=& \rho(\epsilon_{\omega}) 
\frac{1-\epsilon_{\omega}/U + 5( \epsilon_{\omega}^2-1)/(4U^2)}{
1-\epsilon_{\omega}/U+3(3\epsilon_{\omega}^2-4)/(4U^2)} \quad , \quad
\left|\omega +\frac{U}{2}-\frac{1}{2U}\right| \leq 2 \; ,
\nonumber 
\end{eqnarray}
where $\epsilon_{\omega}$ is the real solution of the cubic equation
\begin{equation}
\omega +\frac{U}{2} + \epsilon_{\omega} +\frac{3- \epsilon_{\omega}^2}{2U}
+\frac{3\epsilon_{\omega}(\epsilon_{\omega}^2-4)}{4U^2} =0 \; .
\end{equation}
Therefore, the gap to second order becomes
\begin{equation}
\Delta_2(U)=U-4-\frac{1}{U} = \Delta_1(U)\; , 
\label{Gap2ndorder}
\end{equation}
because $\mu_{\rm LHB}^+=-U/2+2+1/(2U)$ does not change to second order 
in~$1/U$. 
All even orders~$1/U^{2n}$ ($n\geq 1$) in the $1/U$~expansion
of the gap are zero for the one-dimensional Hubbard model\cite{LiebWu},
and the same might be true in infinite dimensions.

Eq.~(\ref{Gap2ndorder}) shows that the Hubbard bands tend to `attract' each
other in the Hubbard model. In contrast to the Falicov--Kimball model,
the gap bends downwards when the interaction is reduced
towards the transition.
We use the large-$U$ expansion to estimate
the value of the critical interaction strength~$U_{\rm c}$
for the Mott--Hubbard transition. From $\Delta_n(U)=0$ ($n=0,1,2$)
in~(\ref{deltazeroHUB}),~(\ref{Gap1storder}),~(\ref{Gap2ndorder})
we find
\begin{equation}
U_{\rm c}^{(0)} = 4 \quad ,\quad
U_{\rm c}^{(1)} = 4.24 \quad ,\quad
U_{\rm c}^{(2)} = 4.24 \quad ,\quad
\ldots \quad . 
\label{UcseriesHUB}
\end{equation}
The shift from $U_{\rm c}^{(0)} =W$ to $U_{\rm c}^{(2)} = 1.06 W$
to second order in~$1/U$ is only six percent. We are presently carrying out
the expansion for the gap to third order in~$1/U$ but we do not expect
sizable corrections, i.e., $U_{\rm c}=(1.05\pm 0.05)W$
is our prediction of the critical interaction strength for the closure
of the gap. 

\begin{figure}[ht]
\centerline{\includegraphics[height=9cm,width=12cm]{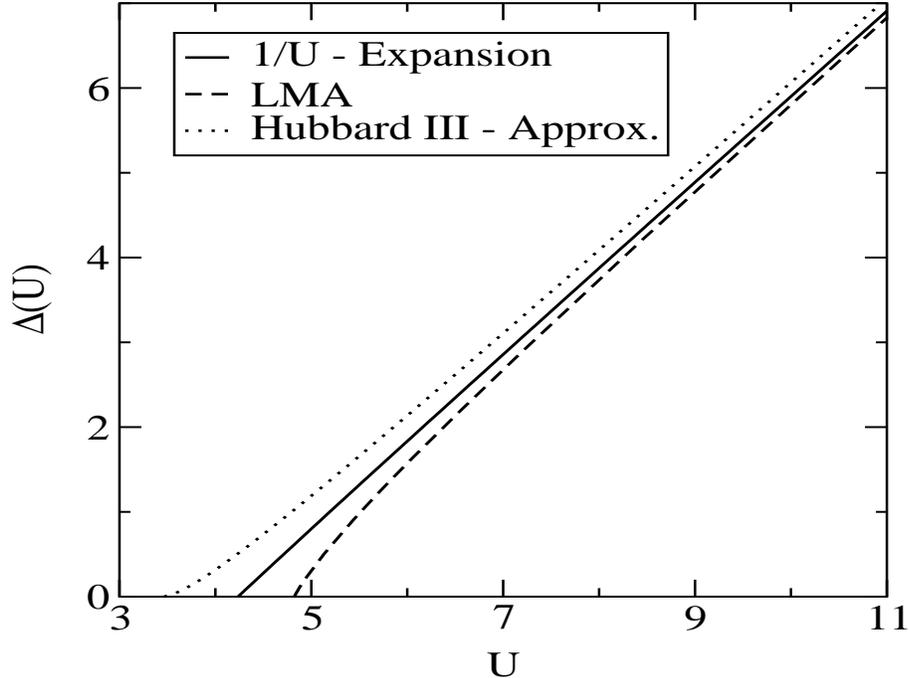}}
%
%\vspace*{4cm}
\caption{Mott--Hubbard gap for one-particle excitations in the
Hubbard model. The result from the expansion to second order 
in~$1/U$ (solid line), 
from the Hubbard-III approximation (dotted line), and
from the local-moment approximation (dashed line) 
are shown.}  
\label{Fig:LMAgap}
\end{figure}

\section{COMPARISON WITH OTHER ANALYTICAL APPROACHES}
\label{SecVI}
\nopagebreak[3]
\subsection{Hubbard-III Approximation}

Hubbard\cite{HubbardIII} provided an early
approximation for the Mott--Hubbard 
insulator. Its simplest form, the `alloy-analogy approximation',
is equivalent to the exact solution of the Falicov--Kimball
model~(\ref{exactGFFK}). The Green function in the (full) Hubbard-III
approximation includes the `resonance broadening corrections'
which lead to the following cubic equation for the Green function
\begin{equation}
3[G_{\rm H-III}(\omega)]^3 -8\omega[G_{\rm H-III}(\omega)]^2 
+4G_{\rm H-III}(\omega)(3+\omega^2-U^2/4) -8\omega=0 \; .
\label{Hubbardthree}
\end{equation}
This approximation becomes exact to $(1/U)^0$, i.e.,
$G_{\rm H-III}(z)\to G_0(z)$ for $U\to\infty$, cf.~(\ref{Gzerodef}).
However, the Hubbard-III approximation predicts 
the transition to occur at $U_{\rm c}^{\rm H-III}
=2\sqrt{3}=3.46=0.866W$. Although this is an improvement over
the alloy-analogy approximation, the Hubbard bands still tend
to repel each other, in contrast to our analytical result~(\ref{UcseriesHUB})
to~${\cal O}(U^{-2})$.

\begin{figure}[ht]
%
%\vspace*{-0.5cm}
\centerline{\includegraphics[height=8.5cm,width=12cm]{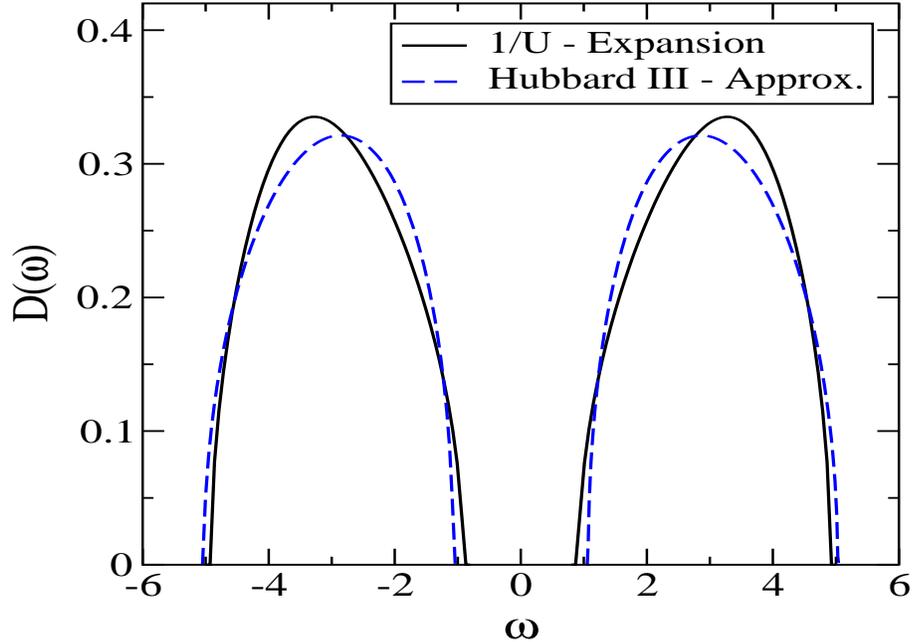}}
%
%\vspace*{-0.5cm}
\caption{Density of states in the
Hubbard model for $U=6$. 
The result from the expansion to second order 
(solid line) and from 
the Hubbard-III approximation (dashed line) are shown.}  
\label{Fig:DOSforHIII}
\end{figure}

The one-particle gap in the Hubbard model as a function of~$U$ is shown in 
Fig.~\ref{Fig:LMAgap} for the Hubbard-III approximation and for the
$1/U$~expansion.
In the Hubbard-III approximation, the gap
has a positive correction in $\Delta_1(U)$,
\begin{equation}
\Delta_1^{\rm H-III}(U)=U-4+\frac{1}{2U} \; , 
\end{equation}
in contrast to the exact first-order result~(\ref{Gap1storder}).
The Hubbard-III approximation overestimates the stability
of the Mott--Hubbard insulator already to first order in~$1/U$ so that
$U_{\rm c}^{\rm H-III}<U_{\rm c}$ appears to be natural.

In Fig.~\ref{Fig:DOSforHIII} we compare the density of states for 
the Hubbard-III
approximation and the $1/U$~expansion for $U=6=(3/2)W$. It is 
seen that the Hubbard-III approximation not only underestimates the gap,
it also fails to reproduce qualitatively the tilt in the
overall structure of the density of states.

\subsection{Local-Moment Approach}

As a second example we compare the results
of our systematic~$1/U$ expansion to those of the 
local-moment approach\cite{LMA}, which provides an 
excellent description of the dynamical properties of
the single-impurity Anderson model, particularly for strong 
coupling\cite{SIAMDavid}.

The gap in LMA acquires the
values\cite{LMA,thankstoDavid}
\begin{equation}
\Delta^{\rm LMA}(U = 11.313) = 7.151 \quad , \quad 
\Delta^{\rm LMA}(U = 8.4658) = 4.226 
\end{equation}
so that we may approximate
\begin{equation}
\Delta_2^{\rm LMA}(U)= U-4- \frac{1.25}{U}-\frac{6.6}{U^2} \; .
\label{GapLMA2nd}
\end{equation}
The comparison with~(\ref{Gap2ndorder}) shows that
the first-order correction is very close to the exact first-order
result.

The one-particle gap as a function of~$U$ is shown in Fig.~\ref{Fig:LMAgap}.
The negative second-order contribution in~(\ref{GapLMA2nd})
makes the local-moment gap bend downwards from the second-order
result~(\ref{Gap2ndorder}). Therefore, the local-moment approach
predicts the collapse of the insulator to occur at 
$U_{\rm c}^{\rm LMA}\approx 4.82 \approx 1.21 W$ which is larger than
our second-order prediction~$U_{\rm c}^{(2)}=4.24\approx 1.06 W$ 
in~(\ref{UcseriesHUB}). 
Since the Hubbard-III approximation underestimates $U_{\rm c}$
and the local-moment overestimates $U_{\rm c}$, 
we are confident that the exact critical interaction
strength cannot be much different from~$U_{\rm c}^{(2)}$. 

\begin{figure}[t]
%
%\vspace*{-1.0cm}
\centerline{\includegraphics[height=10.5cm,width=13cm]{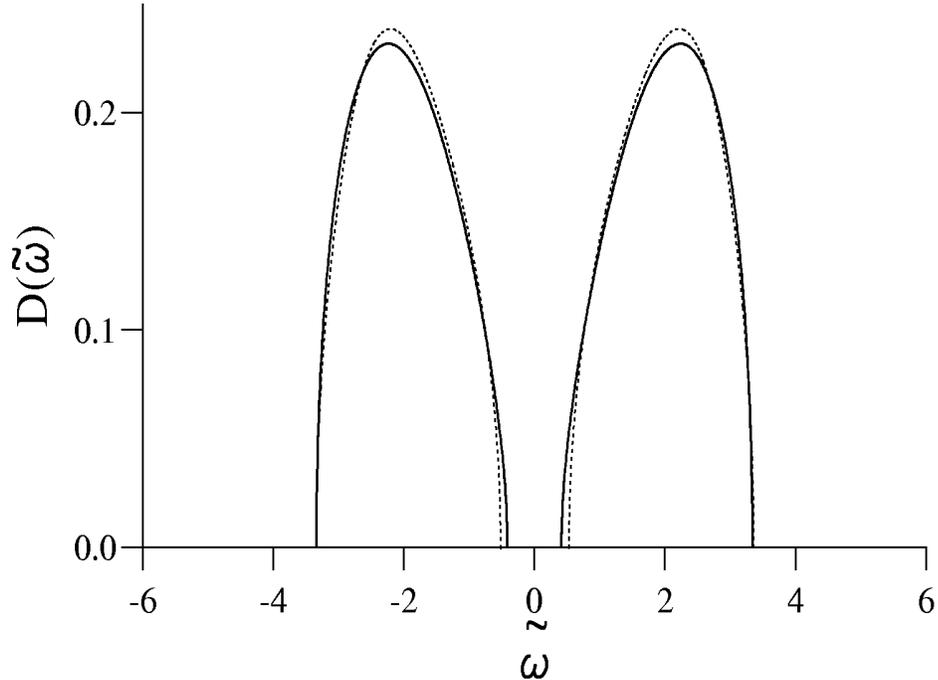}}
%
%\vspace*{-0.2cm}
\caption{Density of states in the
Hubbard model for $U=4\sqrt{2}\approx 5.66$. 
The result from the expansion to second order 
(dashed line) 
and from the local-moment approach (solid line) are shown.}  
\label{Fig:DOSforLMA}
\end{figure}

In Fig.~\ref{Fig:DOSforLMA} we display the density of states
for the local-moment insulator\cite{LMA} for $U=4\sqrt{2}\approx 5.66
\approx 1.4W$
together with the corresponding result~(\ref{DOSHUB2ndorder})
of the $1/U$~expansion to second order. 
Note that the energy scale is $t^*=\sqrt{2}t$ in Fig.~\ref{Fig:DOSforLMA},
and the density of states is rescaled accordingly\cite{LMA}.
It is seen that the overall
agreement is very good. 
The deviations are largest in the vicinity of the band edges, i.e.,
the single-particle gap deviates more than
the overall density of states. 
In fact, the two curves are almost indistinguishable
for $U=6\sqrt{2}\approx 8.49\approx 2.1W$.
This shows that the local-moment approach provides a very reliable
description of the Mott--Hubbard insulator for $U>1.4 W$.

\section{CONCLUSIONS}
\label{SecVII}

In this work we have formulated the $1/U$~expansion for the single-particle
Green function for the Mott--Hubbard insulator on the Bethe lattice
with infinite coordination number. 
For this case the terms to order~$1/U^{n}$ in the Kato--Takahashi 
perturbation theory can be replaced by two $n$th-order polynomials
in $\hat{h}_0$, one for the
shape-correction correction terms and one for
the gap-renormalization terms. The latter polynomial has only even (odd) powers
for $n$~odd (even). 
Therefore, the density of states, which is an important ingredient
for the dynamical mean-field theory, is actually characterized by
a (small) set of numbers for $U\gg U_{\rm c}$. This observation
should be useful for an assessment, and possible improvement, of the quality of
numerical techniques where the continuous density
of states is replaced by a few peaks\cite{RMP,Krauth,BullaNRGPRL,ED1,ED2}.

We have tested our approach against
the exact solution for the Falicov--Kimball model. 
The $1/U$~expansion up to second order gives an excellent
description for $U>3t=1.5 U_{\rm c}^{\rm FK}$. 
The extrapolated gap is only 10\% 
off the exact result.
We are confident that the $1/U$~expansion is equally reliable
for the Hubbard model in the regime $U_{\rm c}/U>2/3$,
where we estimate $U_{\rm c}=(1.05\pm 0.05)W$ from the extrapolation
of the gap to second order in~$1/U$. 

As a first application of our result, we have tested two approximate
theories for the Mott--Hubbard insulator which are exact
to order~$(1/U)^0$. The first-order correction
to the gap in the Hubbard-III approximation resembles that of the
Falicov--Kimball model, i.e., it has the wrong sign. Therefore,
the density of states in the Hubbard-III approximation
qualitatively disagrees with the $1/U$~expansion, and
the value for the Mott--Hubbard transition, $U_{\rm c}^{\rm H-III}=2\sqrt{3} W
\approx 0.866W$ is too small.

The density of states
from the local-moment approach agrees very well with our results down
to $U\approx 1.4 W$. The first-order corrections to the gap 
are very close but the local-moment approach has a sizable
second-order correction which is actually absent in the Hubbard model.
Therefore, the local-moment estimate for the Mott--Hubbard transition 
is somewhat too large, $U_{\rm c}^{\rm LMA}\approx 1.21 W$.

There are two applications of our results.
First, our formulae can be used as an input guess for the 
iteration cycle of the dynamical mean-field theory.
This may help to stabilize and speed up the convergence of the
iteration in the Mott--Hubbard insulator for all $U>U_{\rm c}$. 
Second, and more importantly,
approximate theories on the Mott--Hubbard {\sl transition\/}
have to pass the test against our 
results for the single-particle density of states 
in the Mott--Hubbard {\sl insulator}.
A more detailed comparison of the existing numerical
and analytical approximations is planned to be published elsewhere.

\section*{ACKNOWLEDGMENTS}

We thank David Logan for sending us the {\sc Postscript} 
originals of Fig.~10 in 
Ref.~\onlinecite{LMA}, and Reinhard Noack for helping us
manipulating the figures.
This work was supported by the Deutsche Forschungsgemeinschaft 
under grant number GE 746/5-1+2 and the Graduiertenkolleg
{\sl Optoelectronics of Mesoscopic Semiconductors}.


\begin{thebibliography}{99}
\bibitem{PvDwarning} The sign of~$t$ in Ref.~\onlinecite{PvD} 
is opposite to our definition which can be cured by another
particle-hole transformation.
\end{thebibliography}
\end{document}